\documentclass[conference]{IEEEtran}
\IEEEoverridecommandlockouts
\usepackage{cite}
\usepackage{amsmath,amssymb,amsfonts}

\usepackage{algorithmic}
\usepackage{graphicx}
\usepackage{textcomp}
\usepackage{xcolor}
\usepackage{balance}

\usepackage{breqn}
\usepackage[font=footnotesize,skip=2pt]{caption}
\ifCLASSOPTIONcompsoc
    \usepackage[caption=false, font=normalsize, labelfont=sf, textfont=sf]{subfig}
\else
\usepackage[caption=false, font=footnotesize]{subfig}
\fi

\def\BibTeX{{\rm B\kern-.05em{\sc i\kern-.025em b}\kern-.08em
    T\kern-.1667em\lower.7ex\hbox{E}\kern-.125emX}}
\begin{document}

\title{Tiny-VBF: Resource-Efficient Vision Transformer based Lightweight Beamformer for Ultrasound Single-Angle Plane Wave Imaging\vspace{-10pt}}

\author{\IEEEauthorblockN{ Abdul Rahoof$^1$, Vivek Chaturvedi$^1$, Mahesh Raveendranatha Panicker$^1$, Muhammad Shafique$^2$}
\IEEEauthorblockA{ 
\textit{$^1$Indian Institute of Technology Palakkad}, India; \textit{$^2$New York University}, Abu Dhabi, UAE}
}

\maketitle

\begin{abstract}
Accelerating compute intensive non-real-time beamforming algorithms in ultrasound imaging using deep learning architectures has been gaining momentum in the recent past. Nonetheless, the complexity of the state-of-the-art deep learning techniques poses challenges for deployment on resource-constrained edge devices. In this work, we propose a novel vision transformer based tiny beamformer (Tiny-VBF), which works on the raw radio-frequency channel data acquired through single-angle plane wave insonification. The output of our Tiny-VBF provides fast envelope detection requiring very low frame rate, i.e. 0.34 GOPs/Frame for a frame size of 368 x 128 in comparison to the state-of-the-art deep learning models. It also exhibited an $8\%$ increase in contrast and gains of $5\%$ and $33\%$ in axial and lateral resolution respectively when compared to Tiny-CNN on in-vitro dataset. Additionally, our model showed a $4.2\%$ increase in contrast and gains of $4\%$ and $20\%$ in axial and lateral resolution respectively when compared against conventional Delay-and-Sum (DAS) beamformer. We further propose an accelerator architecture and implement our Tiny-VBF model on a Zynq UltraScale+ MPSoC ZCU104 FPGA using a hybrid quantization scheme with $50\%$ less resource consumption  compared to the floating-point implementation, while preserving the image quality.
\end{abstract}


\section{Introduction}
\par Ultrasound (US) imaging is an important diagnostic method which is used extensively due to its cost-effectiveness, convenience and absence of ionizing radiation. Conventional ultrasound imaging employs a focused transmit scheme that systematically reconstructs the region of interest line by line. However, this method's frame rate is restricted, making it unsuitable for scenarios involving significant motion in the area being scanned, as seen in echocardiography, elastography, and doppler applications~\cite{pw}. On the other hand, plane wave (unfocused) transmit schemes are popular for their higher frame rates, typically exceeding a few 1000 Frames/sec~\cite{pw}, but faces a significant challenge in achieving satisfactory image quality, particularly resolution and contrast, due to its complex beamforming process during image reconstruction~\cite{pw}.

\par Ultrasound imaging systems commonly adopt the Delay-and-Sum (DAS)~\cite{das} beamforming technique due to its lower computational demands. However, when combined with single plane wave imaging, DAS leads to reduced resolution and contrast due to the data-independent apodization process. Coherent Plane-Wave Compounding offers an adept solution to mitigate the reduction in image quality~\cite{cpw}. However, this solution comes with the drawback of decreased frame rate, resulting in a trade-off between image quality and frame rate. A significant advancement in this realm is the Minimum Variance Distortionless Response (MVDR)~\cite{mvdr}.  However, the computational complexity of MVDR, particularly the matrix inversions poses challenges for real-time implementation having a complexity of $O(n^3)$, where $n$ is the number of channels. For instance, it demands approximately 98.78 GOPs/Frame for a frame size of 368 x 128~\cite{mvdr2}. 
\par Recently, several works have shown significant performance improvement in image reconstruction by using deep learning models for beamforming process~\cite{fcnn, cnnb, unet,googlenet,mobilenet}. The integration of deep learning models enhances the quality of the reconstructed images with single angle plain wave insonification. However, the complexity of these models such as~\cite{unet} (50 GOPs/Frame) and~\cite{googlenet} (199 GOPs/Frame) makes them non-trivial to implement on resource-constrained devices. 

Evidently, compact ultrasound imaging system holds the potential to enhance diagnostic confidence, improve patient care and offer the advantages of portability and accessibility in medical settings. These compact systems or edge devices have stringent resource constraints and achieving high performance on such systems is non-trivial. Employing Field Programmable Gate Arrays (FPGAs) in beamforming emerges as a compelling choice due to their flexibility and parallel processing capabilities, making them a more efficient solution compared to Graphics Processing Units (GPUs) and processors like Media Processors (MP) and Digital Signal Processors (DSP) for enhancing ultrasound imaging systems, particularly in portable and resource-constrained environments~\cite{compact}. 

\par Towards the above mentioned objective, we propose a vision transformer (ViT)~\cite{attention} based ultrasound image reconstruction beamformer Tiny-VBF which is accelerated on a FPGA based hardware evaluation board. Recently, ViTs have gained immense popularity, revolutionizing the accuracy of pivotal computer vision tasks. The calculation pattern employed in ViTs differs from the convolutional units used in convolutional neural networks (CNNs) for extracting partial information. ViTs prioritize the comparison of pixel channels rather than focusing solely on feature values within the receptive field. It employs stacked self-attention and point-wise fully connected layers in both the encoder and decoder. The self-attention layers in ViTs are global in nature and establish comprehensive dependencies between input and output. Currently, ViT models are rapidly surpassing CNN models in various computer vision tasks~\cite{swin}. Next, we discuss the major contributions of our work.
\par
\begin{figure}[h]
    \centering
  
  \subfloat[\label{1a}]{%
        \includegraphics[scale=0.11]{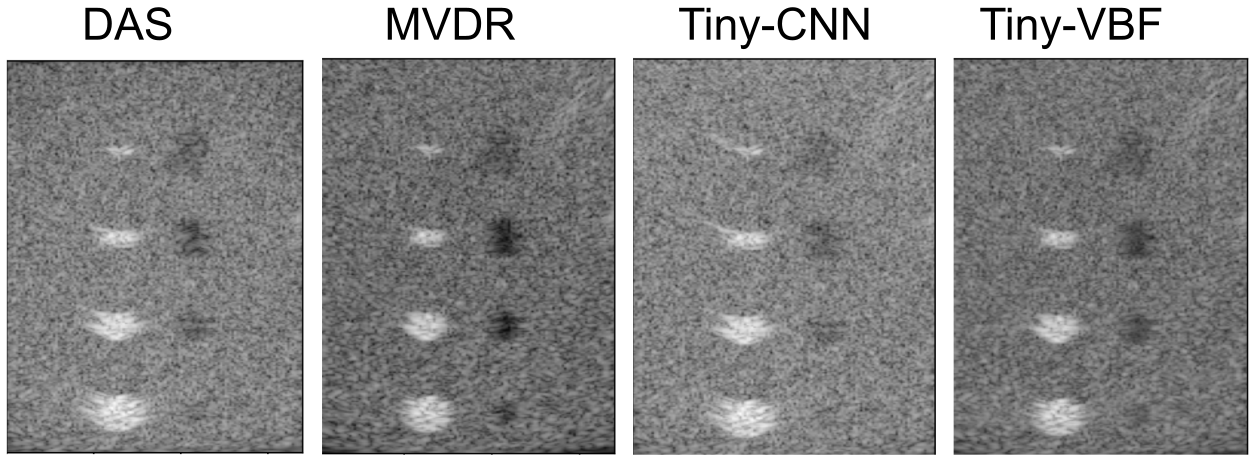}}
  \hspace{0.5mm}
  \subfloat[\label{1b}]{%
        \includegraphics[scale=0.06]{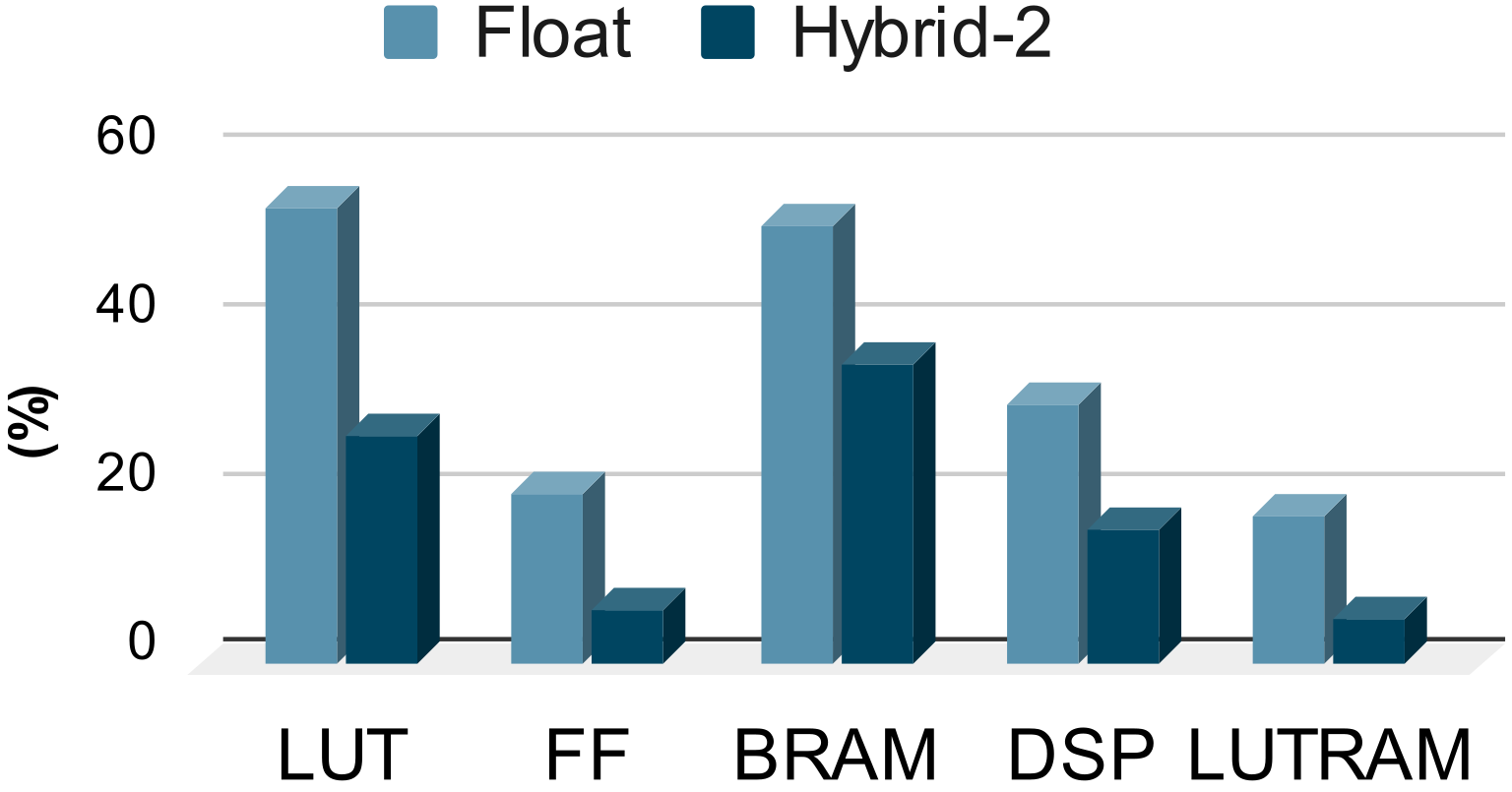}}
  \caption{(a) The B-mode image of cyst evaluation experiment data; and \hspace{1cm}(b) Comparison of resource consumption of Float and Hybrid-quantized Tiny-VBF model on FPGA}
  \label{fig:my_label-00} 
\end{figure}

 \textbf{Our Novel Contributions:} 
\vspace{-1mm}
\begin{itemize}
    \item We propose a ViT-based image reconstruction model, Tiny-VBF, which improves the quality of ultrasound images, achieves more accurate visualization of structures and abnormalities compared to conventional DAS beamformer. Moreover, Tiny-VBF can effectively detect edges compared to DAS and Tiny-CNN~\cite{cnnb} beamformers, as illustrated in the reconstructed images presented in Fig.~\ref{fig:my_label-00}(a).
    \item Our Tiny-VBF is computationally light and uses fewer FLOPs (0.34 GOPs/Frame) compared to the state-of-the-art tinyML models Tiny-CNN (11.7 GOPs/Frame) and FCNN~\cite{fcnn} (1.4GOPs/Frame).
    \item We performed an extensive quality evaluation in terms of contrast and resolution on two differnt datasets, i.e. in-vitro dataset and in-silico dataset.
    \item For in-vitro data, the Tiny-VBF showed an $8\%$ increase in contrast, while also achieving gains of $5\%$ and $33\%$ in axial and lateral resolution compared to the Tiny-CNN, and a $4.2\%$ increase in contrast with gains of $4\%$ and $20\%$ in axial and lateral resolution compared to the DAS.
    \item Additionally, for in-silico data, the Tiny-VBF showed a $10.7\%$ increase in contrast, while also achieving gains of $17.7\%$ and $25\%$ in axial and lateral resolution compared to Tiny-CNN, and an $8\%$ increase in contrast with gains of $16.8\%$ and $25\%$ in axial and lateral resolution compared to DAS. 
    \item We implemented our Tiny-VBF on the ZCU104 FPGA board (with clock frequency of 100 MHz) and conducted a comprehensive assessment of resource utilization and image quality across quantization, including our hybrid quantization. The hybrid quantization strategy led to a remarkable $50\%$ reduction in resource consumption compared to the floating-point implementation, while maintaining image quality, as illustrated in Fig.~\ref{fig:my_label-00}(b).
\end{itemize}

\vspace{-1mm}

\par
\begin{figure}[h]
\centering
\includegraphics[scale=0.13]{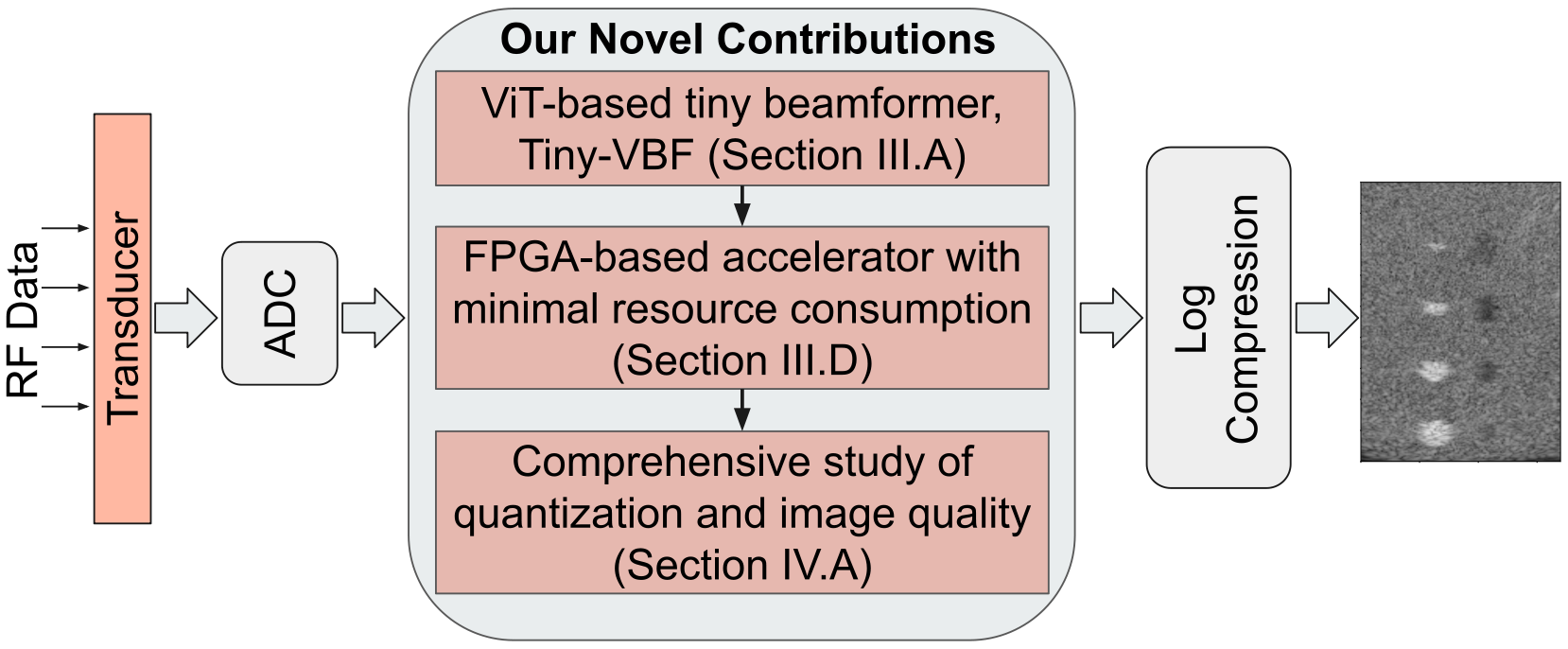}
\caption{Overview of our novel contributions}
\label{fig:my_label-0}
\end{figure}

 To the best of our knowledge, we are the first to evaluate the applicability of beamforming through a vision transformer network and deploy a deep learning-based beamformer on resource-constrained edge devices. In Fig.~\ref{fig:my_label-0}, we present an overview of our innovative approach. 
\par \textbf{Paper Organization:} Section II provides an overview of related works, followed by Section III which outlines the proposed methodology.   Section IV details  the experimental results, and lastly, Section V presents the conclusion.

\section{Related Works}
\vspace{-1.5mm}
\par There are several machine and deep learning approaches to improve various signal processing tasks in ultrasound imaging with emphasis on beamforming. The work~\cite{refart} proposed a preprocessing technique for the detection and elimination of reflection artifacts within photoacoustic channel data. Similarly, the work in~\cite{supoff} proposed a preprocessing technique for suppressing offaxis scattering in the received channel data. In the works of ~\cite{den1} and~\cite{den2}, proposed a deep learning-based denoising method for ultrasound single angle plane wave imaging. 
\par The work described in~\cite{fcnn} designed a fully connected neural network (FCNN) to perform beamforming on a pixel-by-pixel basis. This work improved the resolution and contrast compared to DAS and showed faster compared to MVDR beamformer. However, FCNN only captures the local features in the input data to estimate the apodization weights. The architecture of the Tiny-CNN beamformer~\cite{cnnb} receives data of a region with dimensions $(x, y, c_h)$, where $x, y,$ and $c_h$ represent the axial and lateral coordinates and the number of transducers (channels), respectively. The product of the output layer (the apodization weights) and the input layer (the ToFC data) is summed along the channel axis to produce the beamformed image data. The resulting image is then processed with the Hilbert Transform to obtain the final B-mode image. Compared to DAS and FCNN, this work improved the resolution and contrast in fewer angles of transmission, which enhanced reconstruction rate. However, the quality of images produced through single-angle plane-wave (PW) imaging is not satisfactory, closely resembling the performance achieved by the DAS method. The work by~\cite{unet} introduced an extensive U-Net model with wavelet decomposition and reconstruction. Notably, this approach exhibited enhanced resolution and contrast in single-angle images when tested on simulation data, showing improvements over the DAS method. Specifically, the axial resolution experienced a remarkable $37.5\%$ enhancement, while the lateral resolution improved by $65.9\%$ compared to DAS. Moreover, the Contrast-to-Noise Ratio (CNR) and Generalized Contrast-to-Noise Ratio (GCNR) were notably elevated by $160\%$ and $33.8\%$, respectively, in comparison to DAS. However, challenges emerged when applying the model to in-vivo data, as non-linear attenuation with depth led to a loss of contrast and vital medium-related information. Additionally, the method exhibited high computational complexity, exceeding 50 GOPs/Frame for a frame size of 368 x 128. Lastly, the approach is non-trivial to implement on resource-constrained edge devices. Similarly,~\cite{googlenet} introduced an innovative approach involving a fusion of the GoogLeNet and U-Net models for beamforming purposes. This method achieved notable improvements in resolution and contrast for both in-vitro and in-vivo data when contrasted with the DAS method. Moreover, the method showed substantial reduction in sidelobes and artifacts, effectively outperforming DAS. Point-reflector in-vitro data revealed a $23.6\%$ enhancement in both axial and lateral resolutions compared to DAS. The Contrast-to-Noise Ratio (CNR) for anechoic in-vitro data achieved $158\%$ enhancement when compared to DAS. However, it is essential to note the high computational complexity of this method, reaching approximately 199 GOPs/Frame for a frame size of 384 x 256. To the best of our knowledge, there have been limited attempts to implement the beamforming accelerators using neural networks on edge devices. In this paper, we have tried to attempt the same using vision transformers.

\section{Proposed Methodology}
\par
\begin{figure}[h]
\centering
\includegraphics[scale=0.14]{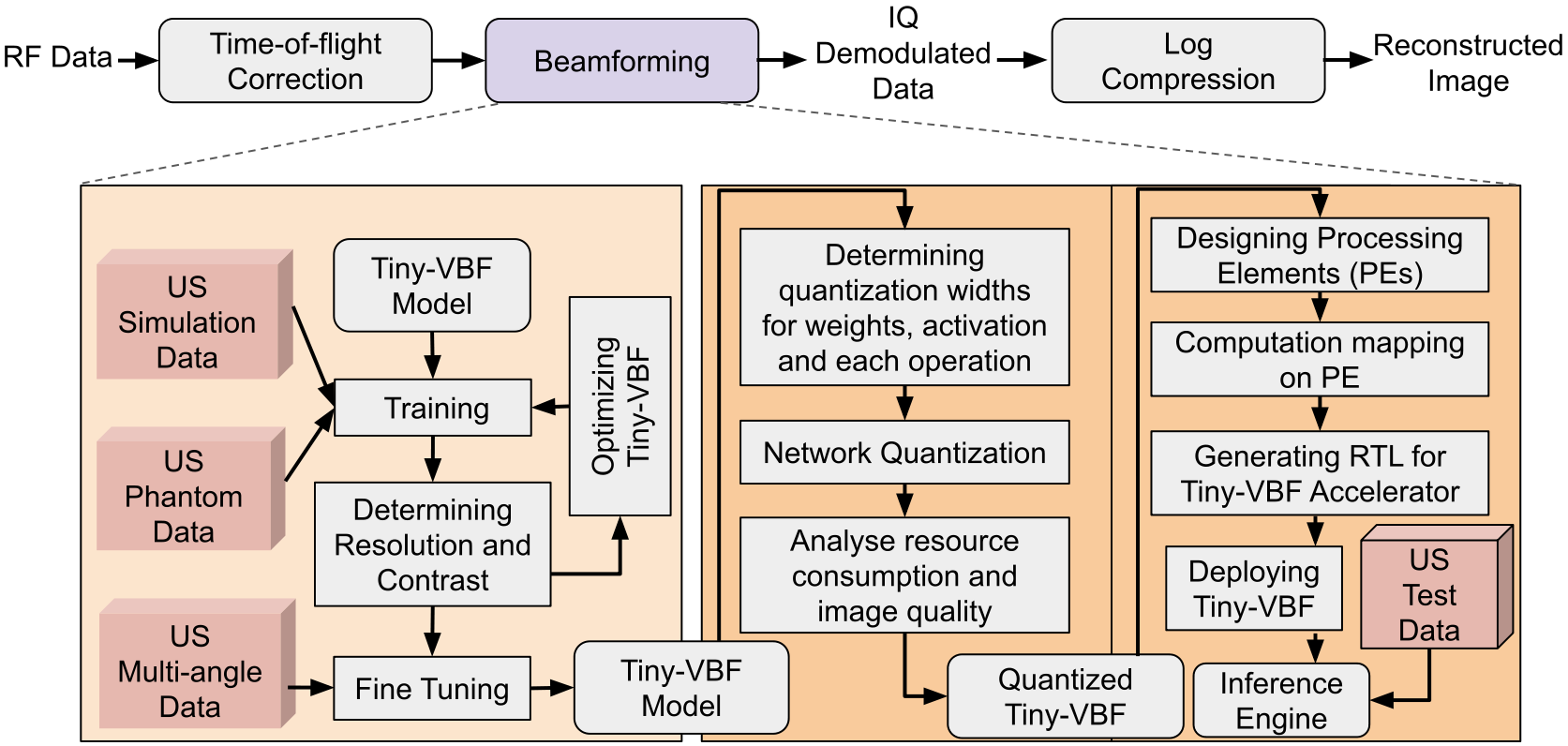}
\caption{Overview of the proposed methodology}
\label{fig:my_label-1}
\end{figure}

 In this section, we describe the implementation of the proposed Tiny-VBF model and its deployment on FPGA. Fig.~\ref{fig:my_label-1} illustrates an overview of the proposed methodology. Our Tiny-VBF Model is crafted through an iterative optimization process, adjusting the encoder-decoder layers to enhance both throughput and resulting image quality. Following this optimization, the model is quantized and an extensive comparison of reconstructed image quality is conducted. Subsequently, an FPGA-based accelerator is designed for deploying the quantized Tiny-VBF model. 

\subsection{Tiny-VBF architecture}
\par
\begin{figure}[h]
\centering
\includegraphics[scale=0.14]{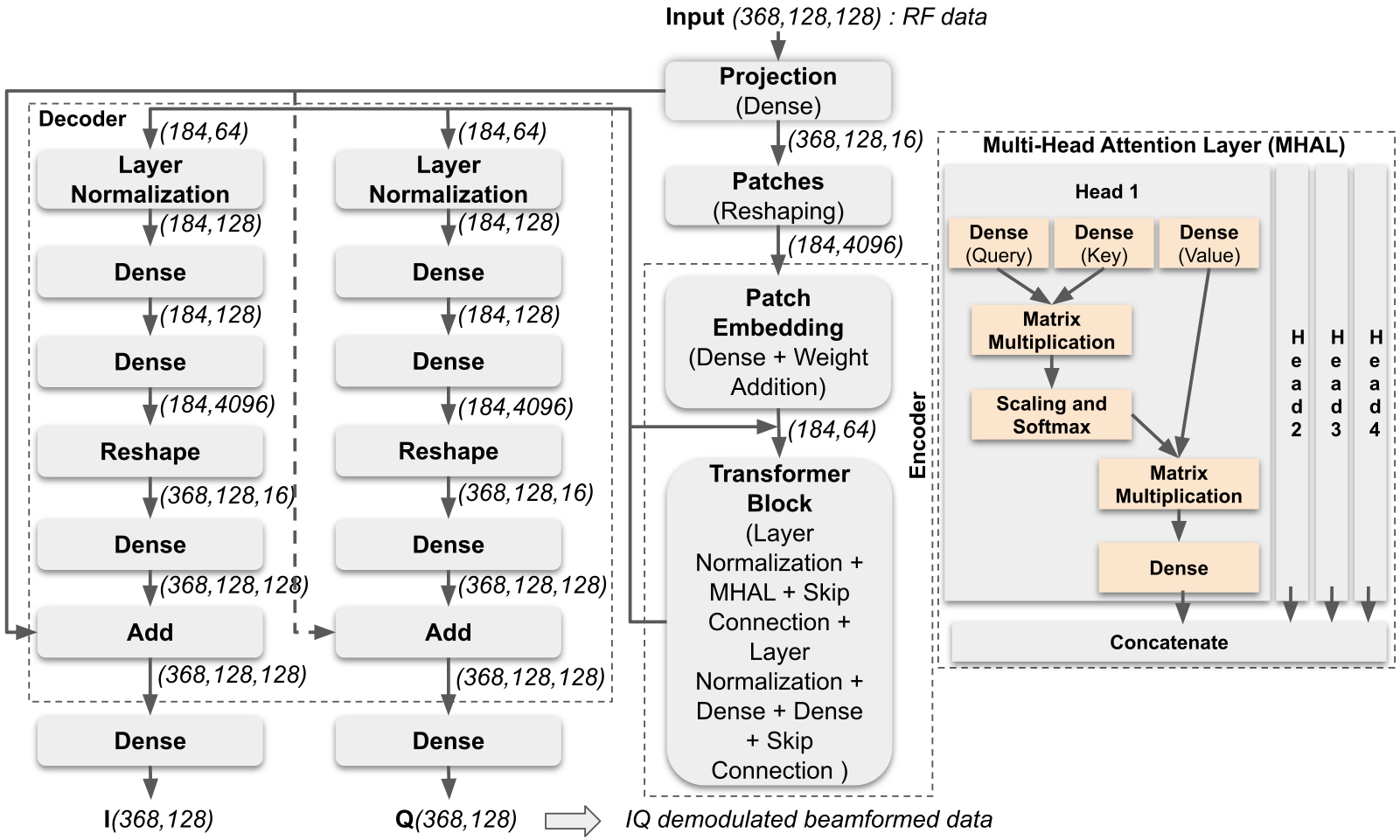}
\caption{The proposed Tiny-VBF architecture}
\label{fig:my_label-2}
\end{figure}

 The architecture of our Tiny-VBF is shown in Fig.~\ref{fig:my_label-2}. The single-angle plain wave data acquired from the ultrasound imaging system undergoes initial time-of-flight correction  before being fed into the Tiny-VBF model. The output of the model is an IQ demodulated beamformed image. The ground truth for learning is the MVDR beamformed IQ demodulated data. The data points were normalized to the interval of [-1, 1]. The architecture includes an encoder block including two transformer blocks and a decoder block. The encoder block is designed such as to obtain the most important features. It contains dense layers to map the channel data to lower dimensions. The transformer block includes a normalization layer, a Multi-Head Attention Layer (MHAL), two skip connectors and two dense layers. The decoder block contains dense layers to reconstruct the IQ demodulated beamformed image from the encoded features.

\par The proposed Tiny-VBF model is implemented using TensorFlow platform (version 2.40) in Python (version 3.8) and was trained and tested on a system  AMD® Ryzen 7 4700g with radeon graphics × 16 with Ubuntu 20.04. The model has a total of 15,07,922 weights. The model reduced the artifacts in the reconstructed image and improved the resolution and contrast of the image compared to standard DAS beamformer. 
\vspace{-2mm}
\subsection{Dataset}

\par The training data was collected from the Verasonics research ultrasound machine, specifically the Verasonics Vantage 128 and L11-5v transducer models (Verasonics Inc., Redmond, WA)~\cite{data1,data2}. The ultrasound machine operates at a center frequency of 7.6 MHz and a sampling frequency of 31.25 MHz. The details of the acquisition are described in~\cite{cnnb,data1,data2}. The data collection process followed the guidelines outlined in the Helsinki Declaration of 1975, which were revised in 2000. Then, the model was fine tuned using the multi-angled images obtained from the CUBDL dataset with 10 angles of transmissions~\cite{cubdl}. 
\par The evaluation of performance is carried out on in-silico and in-vitro datasets provided by the plane-wave imaging challenge for medical ultrasound (PICMUS) of the 2016 IEEE International Ultrasonics Symposium~\cite{picmus}. The in-silico data for PICMUS is generated using Field II, while the in-silico dataset is acquired using a Verasonics Vantage 256 research ultrasound scanner (Verasonics Inc., Redmond, WA).

\subsection{Parameter Setting}
\par The proposed technique involves several hyperparameters that are crucial for its implementation. These include the batch size, optimizer, learning rate and loss function. The batch size was set to 10, and the optimization was performed using the Adam optimizer. The learning rate was not constant but was subjected to a polynomial decay schedule with cyclic changes, which allowed for more flexible parameter updates. Specifically, the initial and final learning rates were set to $10^{-4}$ and $10^{-6}$, respectively. During each learning stage, the number of epochs was set to 1000. The loss function for this technique is Mean Squared Error (MSE) applied to the IQ demodulated beamformed image prior to log compression.

\subsection{Tiny-VBF Accelerator}
\par
\begin{figure}[h]
\centering
\includegraphics[scale=0.11]{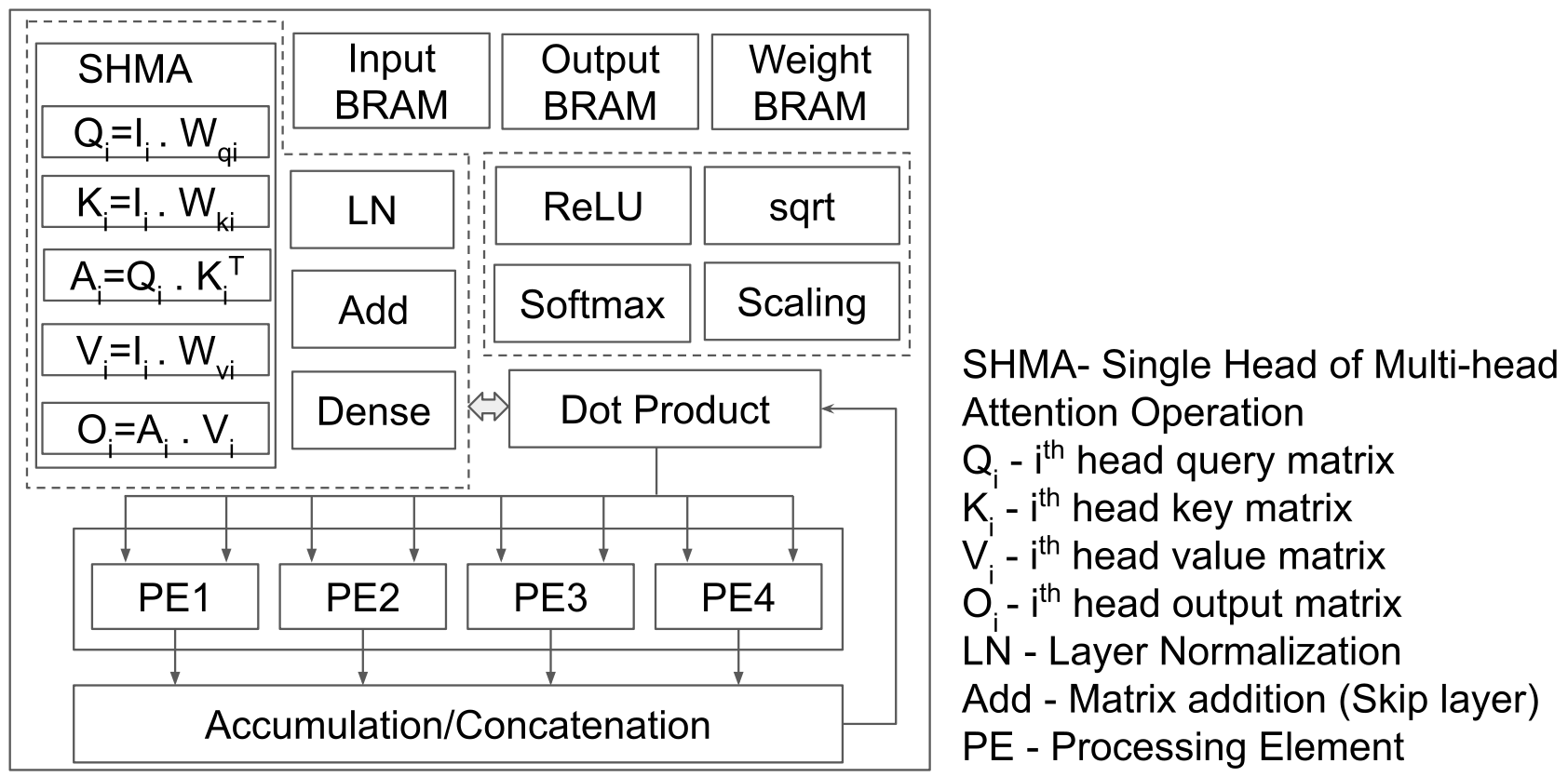}
\caption{The proposed Tiny-VBF accelerator architecture}
\label{fig:my_label-3}
\end{figure}

 The accelerator architecture of the Encoder block is shown in Fig.~\ref{fig:my_label-3}. The architecture involves following non-linear operations ReLU, softmax, division and sqrt. It has 4 Processing Elements (PEs), each of which operates 16 element-wise multiplications followed by an adder tree. The input, weights and intermediate results of the operations are stored in on-chip BRAM memory. 

\par
\begin{figure}[h]
\centering
\includegraphics[scale=0.11]{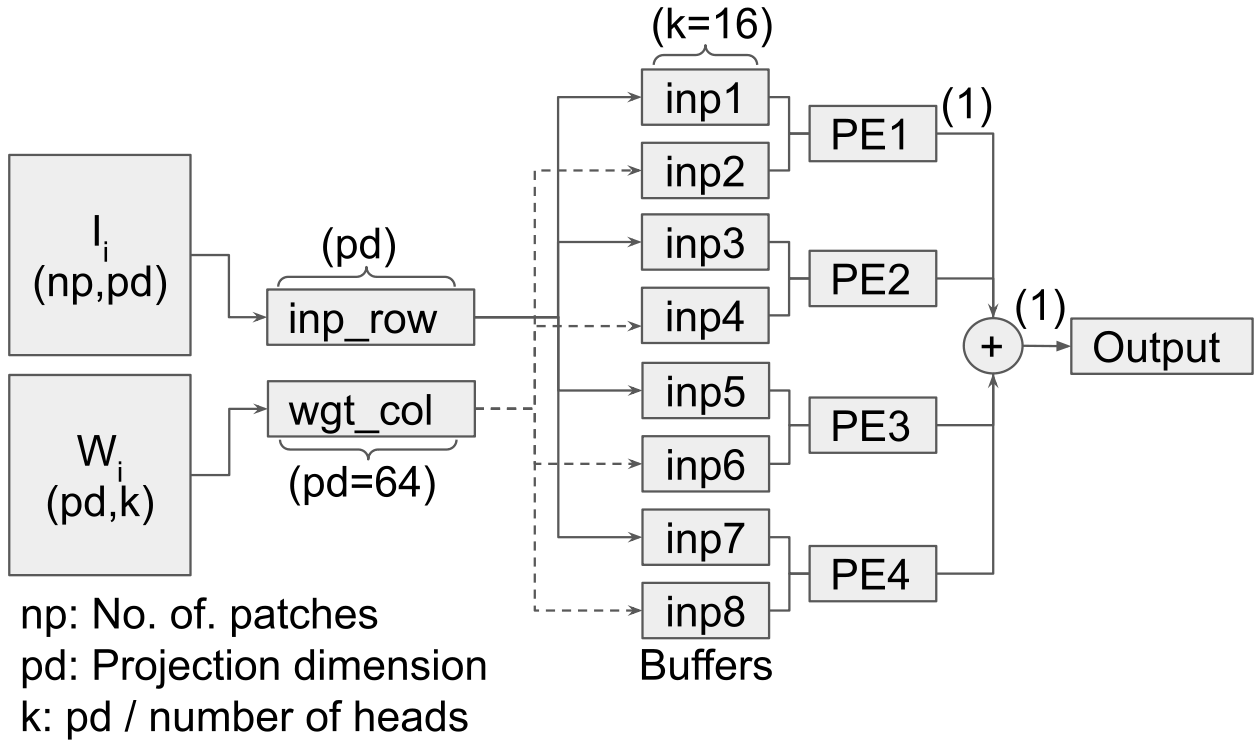}
\caption{Operation involved in computing Query, Key and Value of MHAL}
\label{fig:my_label-4}
\end{figure}

 The operations involved in computing Query, Key and Value of $i^{th}$ head is shown in Fig.~\ref{fig:my_label-4}. The input $I_i$ will be multiplied by Weight $W_i$ to get the resultant matrix with size $(np, k)$ where, $np$ is the number of patches and $k$ is the projection dimension divided by the number of heads. The operation will be performed on the processing elements $PE1, PE2, PE3$ and $PE4$. Here, the $inp1, inp2,...,inp8$ are loaded with the values of the row and column of input and weight, and perform dot products using the PEs. The output values will be accumulated and stored to the output memory.

\par
\begin{figure}[h]
\centering
\includegraphics[scale=0.11]{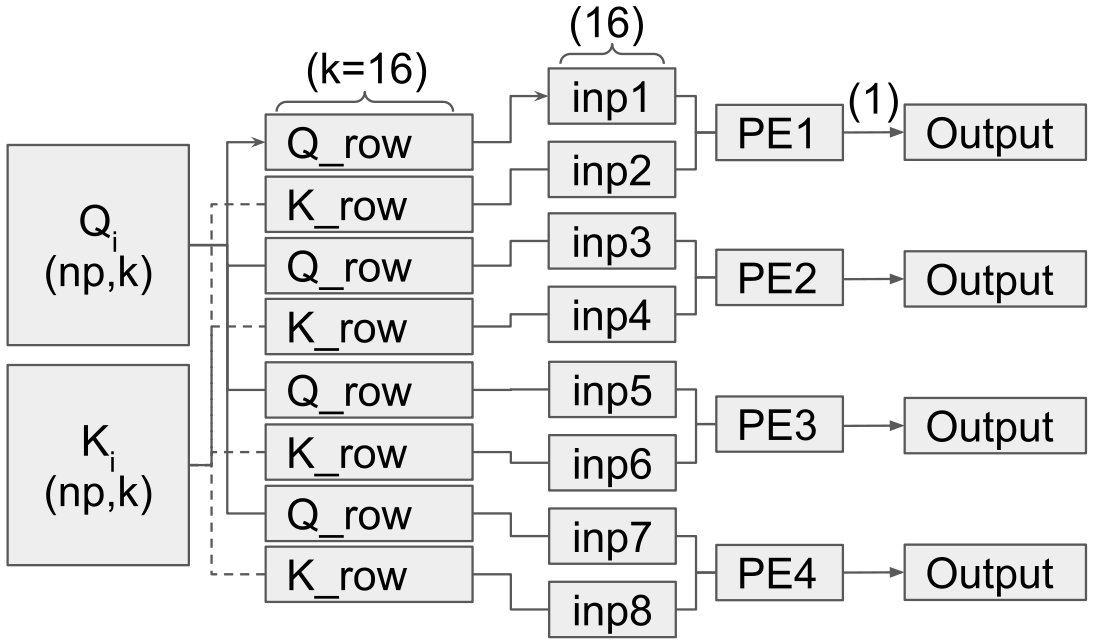}
\caption{Architecture for computing the attention score of single head operation}
\label{fig:my_label-5}
\end{figure}

\begin{figure}[h]
    \centering
  
  \subfloat[\label{1a}]{%
        \includegraphics[scale=0.11]{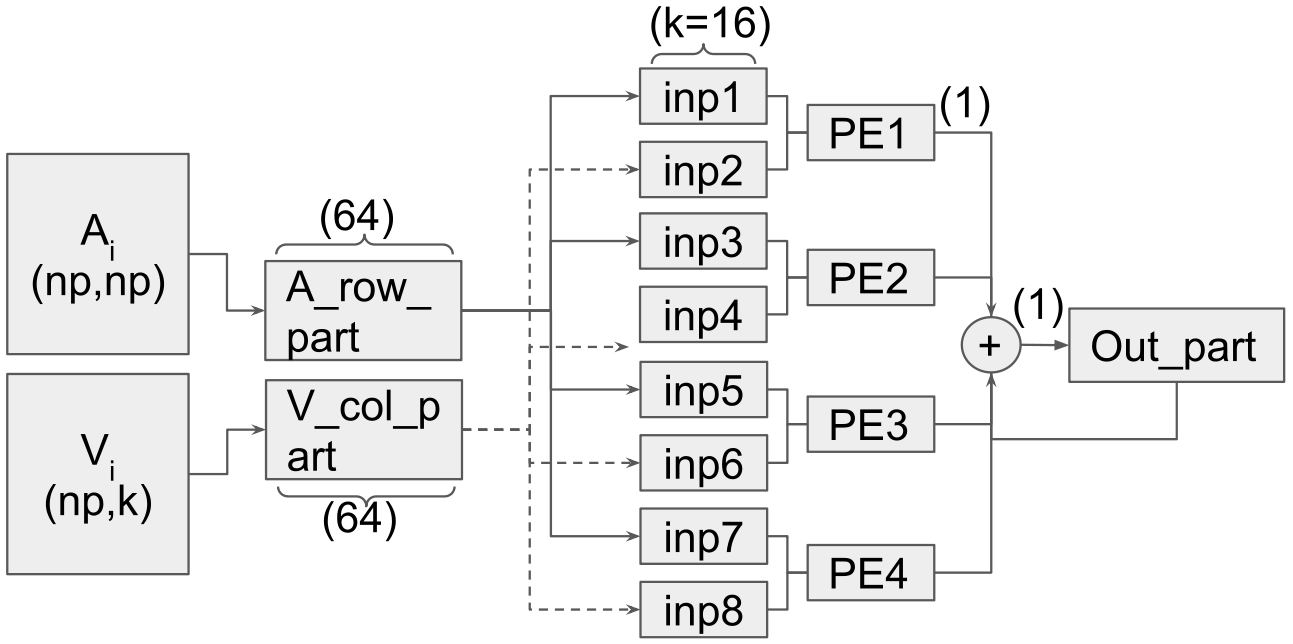}}
  \hspace{0.5mm}
  \subfloat[\label{1b}]{%
        \includegraphics[scale=0.1]{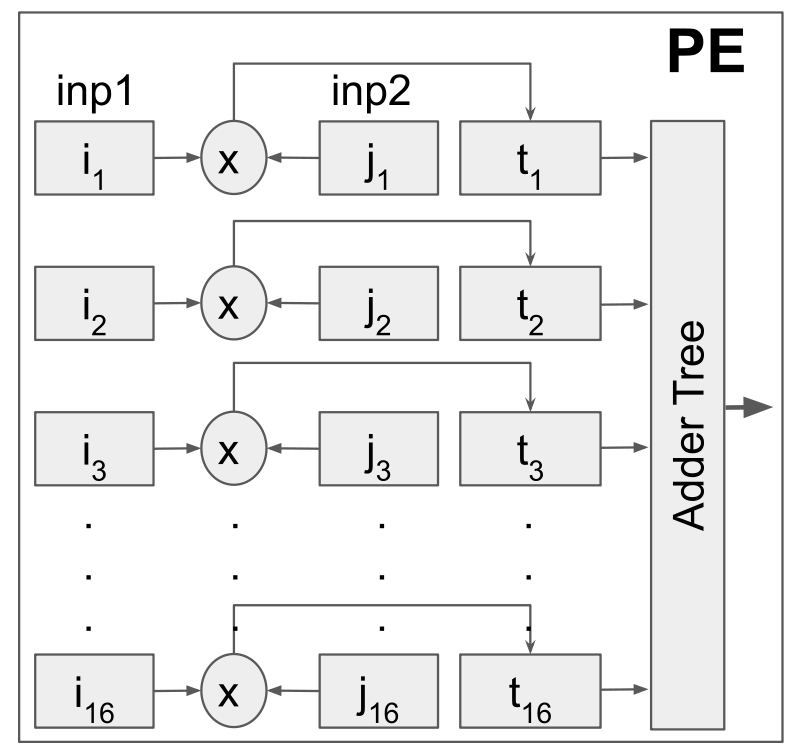}}
  \caption{(a) Architecture for computing the output of single head operation or dense layer operation and  (b) Architecture of single Processing Element}
  \label{fig:my_label-6} 
\end{figure}

 Similarly, the attention score for each head will be computed as shown in Fig.~\ref{fig:my_label-5}. The Query and Key matrices have dimension $(np, k)$ and the resulting matrix has a dimension of $(np,np)$. Moreover, the dense operation and the output of the single head operation, that is the result of multiplying attention score and Value matrices, will be computed by using the architecture shown in Fig.~\ref{fig:my_label-6}(a). The partial result needs to be accumulated with the next corresponding row and column operation, resulting in the final output, which is then stored in the output memory. 
Finally, Fig.~\ref{fig:my_label-6}(b) represents the architecture of the PE. It involves 16 element multiplications followed by an adder tree.

\section{Experimental Results}
\par In this section, a comprehensive examination of the proposed Tiny-VBF is presented, using diverse in-silico and in-vitro datasets that include point targets and cyst targets. 
\par
\begin{figure}[h]
    \centering
  
  \subfloat[\label{1a}]{%
        \includegraphics[scale=0.1]{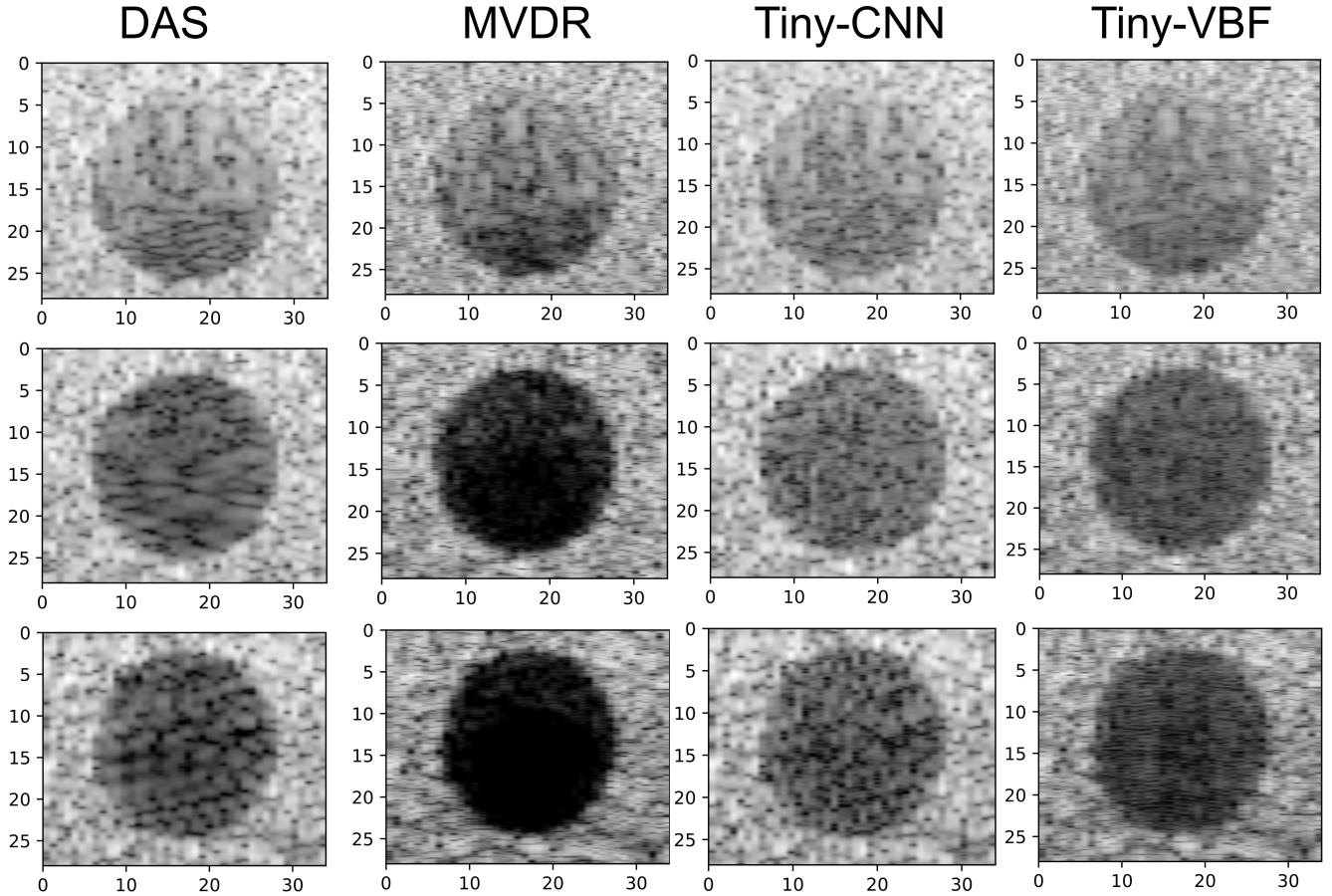}}
  \hspace{0.5mm}
  \subfloat[\label{1b}]{%
        \includegraphics[scale=0.09]{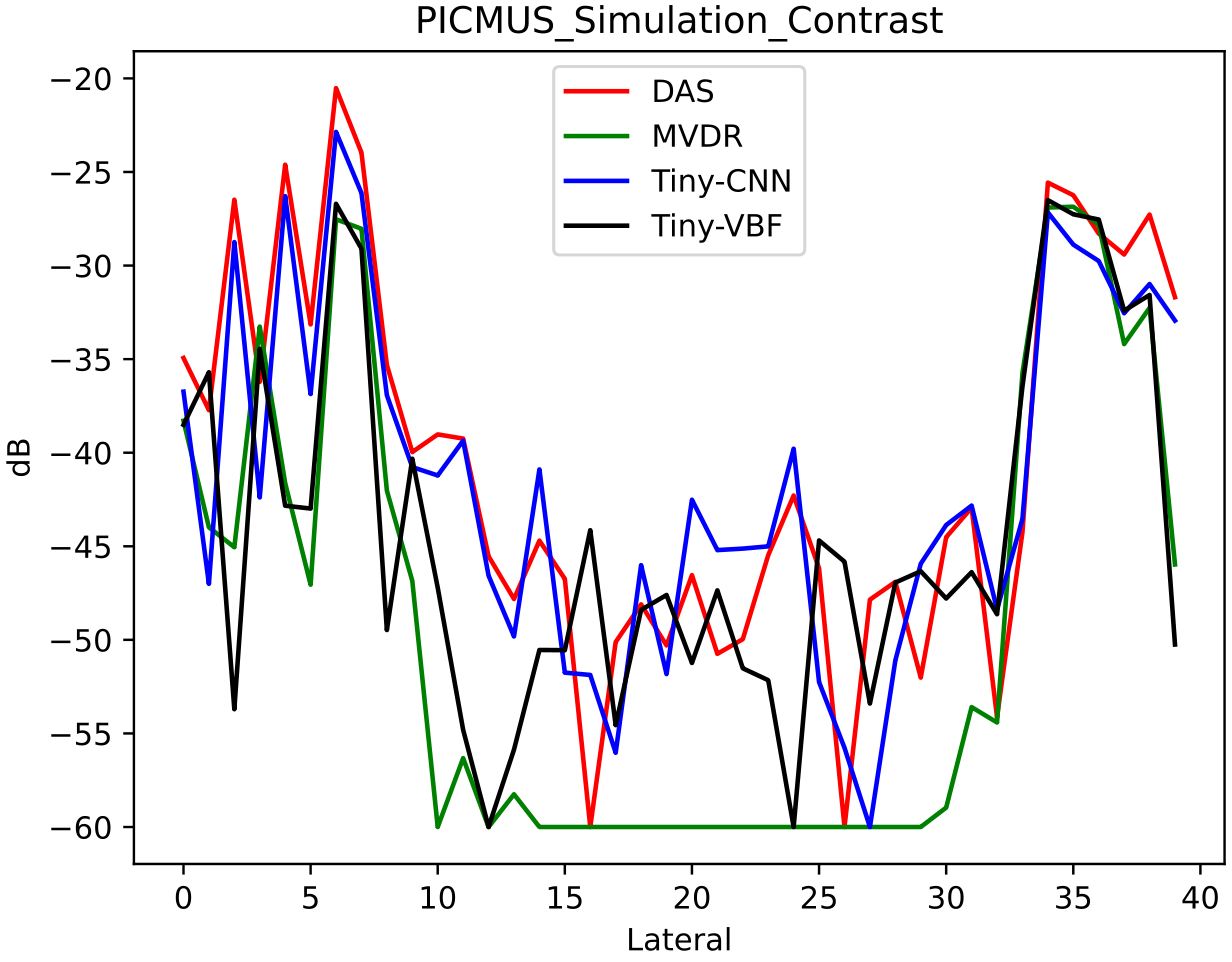}}
  \caption{(a) The in-silico contrast images at three different depths, ranging from near field to far field (at depths of 13 mm, 25 mm, and 37 mm) and  (b) Lateral varitations in the beamformed cyst from the in-silico contrast data at depth of 37 mm}
  \label{fig:my_label-8} 
\end{figure}

 Fig.~\ref{fig:my_label-8}(a) demonstrates the results of each ultrasound beamforming method over single angle PICMUS contrast simulation data. The performance of the state-of-the-art Tiny-CNN model on this data is restricted to that of the DAS. There remains noise and other artifacts in the region of cysts in case of DAS and Tiny-CNN, but it is reduced greatly in case of Tiny-VBF and our benchmark MVDR. From the Fig.~\ref{fig:my_label-8}(b), it is evident that the Tiny-VBF and target MVDR exhibit sharp changes in pixel intensity along the lateral position on the image boundary of cysts, as opposed to the standard DAS and Tiny-CNN.

\begin{figure}[h]
\centering
\includegraphics[scale=0.13]{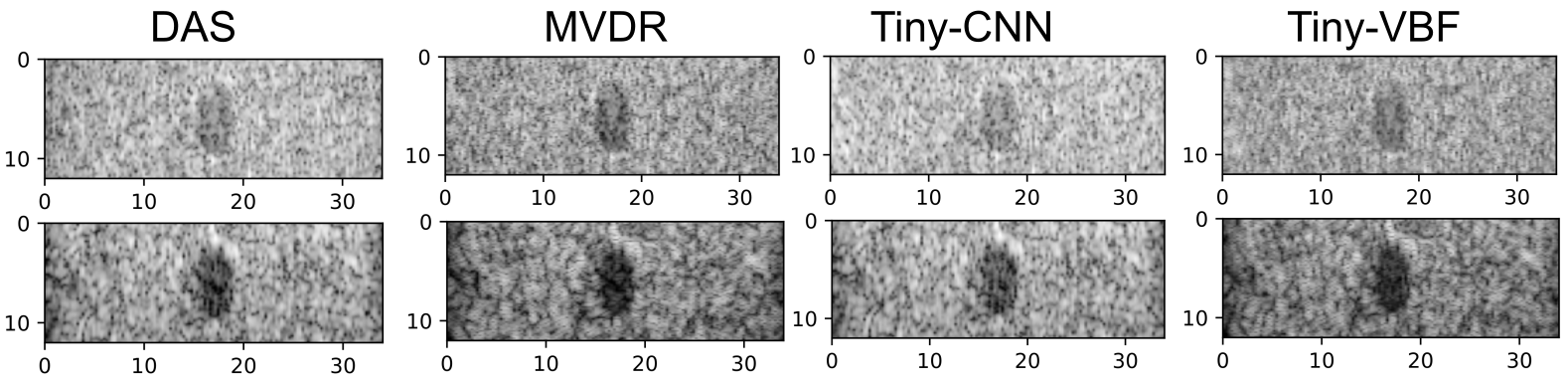}
\caption{The in-vitro contrast images at depths of 15 mm and 35 mm}
\label{fig:my_label-10}
\end{figure}

Similarly, Fig.~\ref{fig:my_label-10} shows the performance of each beamforming method over in-vitro contrast data. The results were obtained from near and far fields (at depths of 15 mm and 35 mm) from the ultrasound probes. It also demonstrates a sharp edge over the cyst region when compared to the DAS and Tiny-CNN.

\par Table~\ref{table-1}  provides a quantitative comparison of the contrast for both simulation (in-silico) and phantom (in-vitro) data. The Contrast Ratio (CR) is higher with Tiny-VBF compared to both DAS and Tiny-CNN, exhibiting improvements of $8\%$ and $10.7\%$ in the in-silico data, and $4.2\%$ and $8\%$ in the in-vitro data, respectively. There is a slight reduction in CNR and GCNR compared to DAS in the simulation data, but it showed better in the phantom data.

\begin{table}[!ht]
        \caption{Contrast metrics (mean) of Simulation and Phantom data}
        \centering
        \begin{tabular}{p{0.18\linewidth}p{0.08\linewidth}p{0.08\linewidth}p{0.08\linewidth}p{0.08\linewidth}p{0.08\linewidth}p{0.08\linewidth}}
            \hline
            \textbf{Beamformer} & \multicolumn{3}{c}{\textbf{Simulation (dB)}} &  \multicolumn{3}{c}{\textbf{Phantom (dB)}} \\
            
            & CR & CNR & GCNR & CR & CNR & GCNR \\
            \hline
            DAS & 13.78 & \textbf{2.37} & \textbf{0.83} & 11.70 & 1.04 & \textbf{0.83} \\
            MVDR & \textbf{21.66} & 1.95 & 0.78 & \textbf{15.09} & \textbf{2.63} & 0.72 \\
            Tiny-CNN & 13.45 & 2.04 & 0.83 & 11.30 & 1.05 & 0.79 \\
            Tiny-VBF & \textbf{14.89} & 1.75 & 0.74 & \textbf{12.20} & \textbf{1.39} & 0.67 \\
            \hline 
        \end{tabular}
\label{table-1}
\end{table}

\par
\begin{figure}[h]
\centering
\includegraphics[scale=0.13]{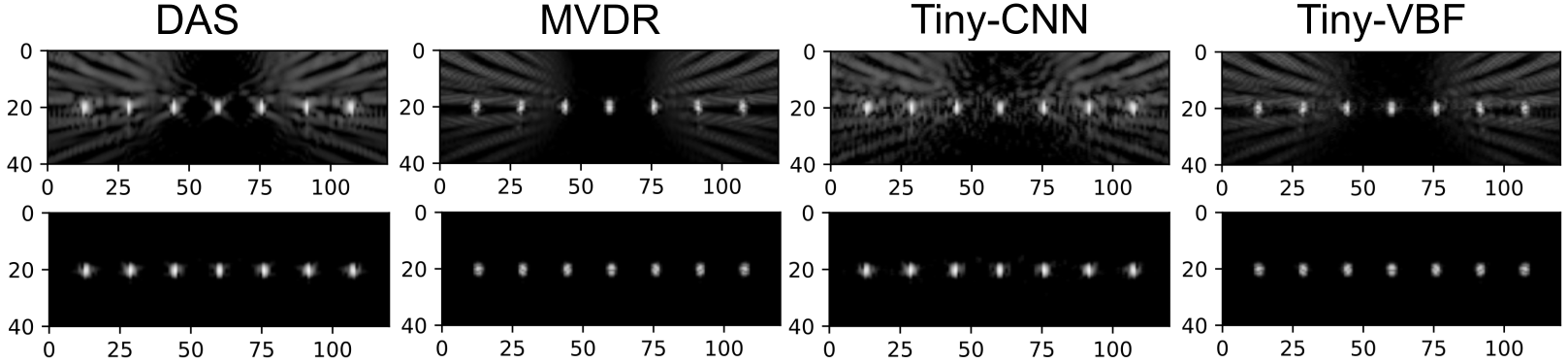}
\caption{B-mode images of in-silico resolution-distortion dataset at depths of 15 mm and 35 mm}
\label{fig:my_label-11}
\end{figure}

 Fig.~\ref{fig:my_label-11} illustrates the results of different beamformers over resolution-distortion simulation data. The point targets are arranged horizontally against an anechoic background in two distinct depth zones. In comparison to Tiny-CNN and DAS, both the Tiny-VBF model and MVDR exhibit superior lateral and axial resolution.

\par
\begin{figure}[h]
    \centering
  
  \subfloat[\label{1a}]{%
        \includegraphics[scale=0.1]{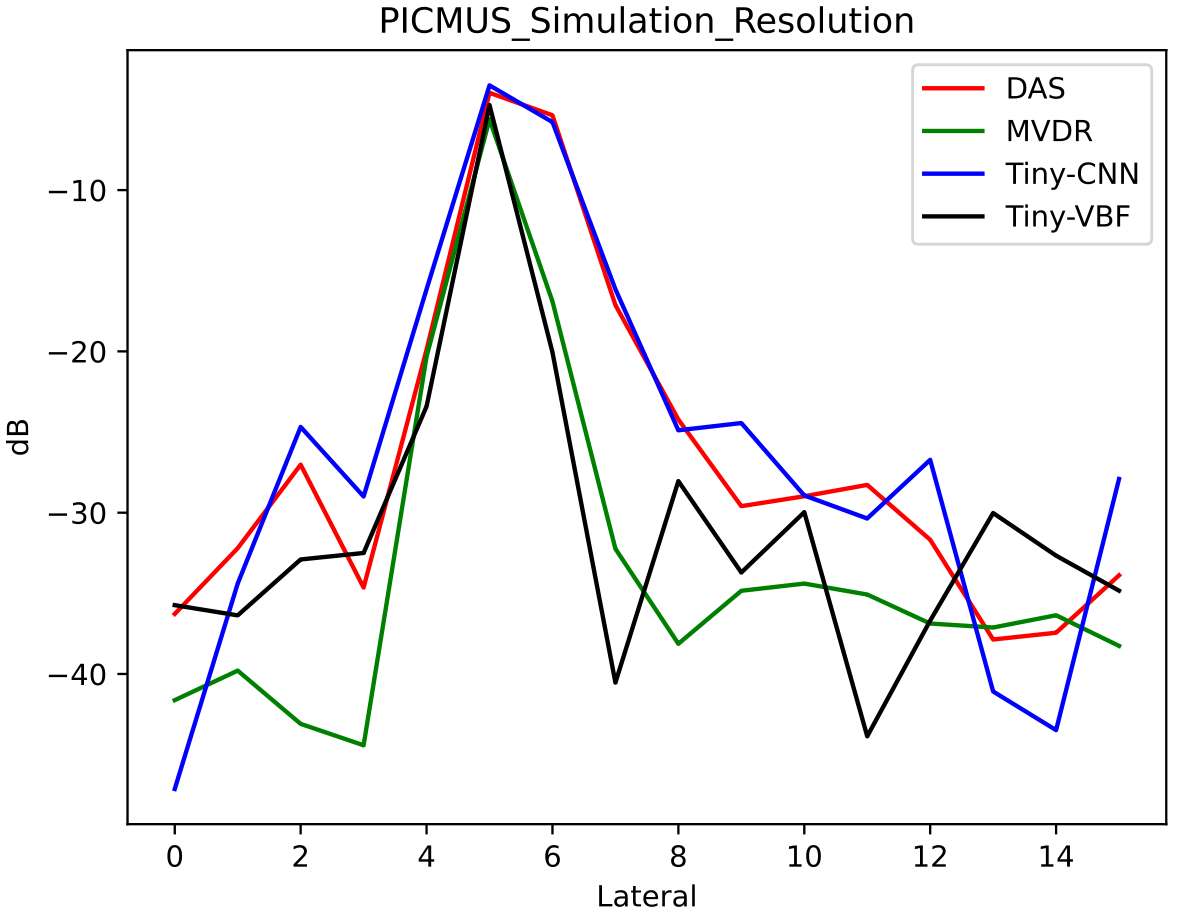}}
        \hspace{0.5mm}
  \subfloat[\label{1b}]{%
        \includegraphics[scale=0.1]{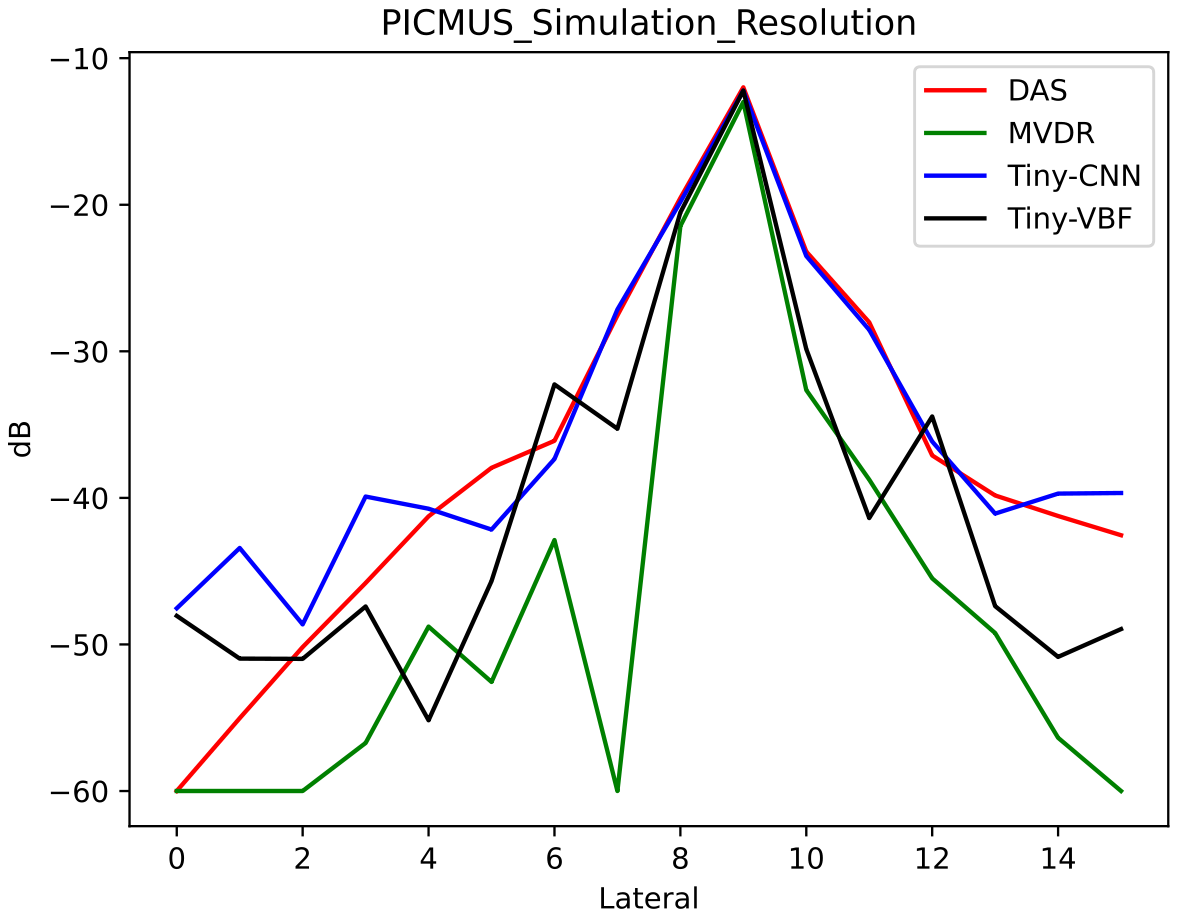}}
  \caption{Lateral point spread functions at (a) 15.12 mm and  (b) 35.15 mm where x-axis represents the lateral point position (pixel position) and y-axis represents the normalized amplitude values}
  \label{fig:my_label-12} 
\end{figure}

 Fig.~\ref{fig:my_label-12} provides a visualization of the lateral resolution for point targets located at  two different depths. It is evident that the MVDR and Tiny-VBF beamformers significantly reduce the mainlobe width and sidelobes in contrast to DAS and Tiny-CNN.

\par
\begin{figure}[h]
\centering
\includegraphics[scale=0.13]{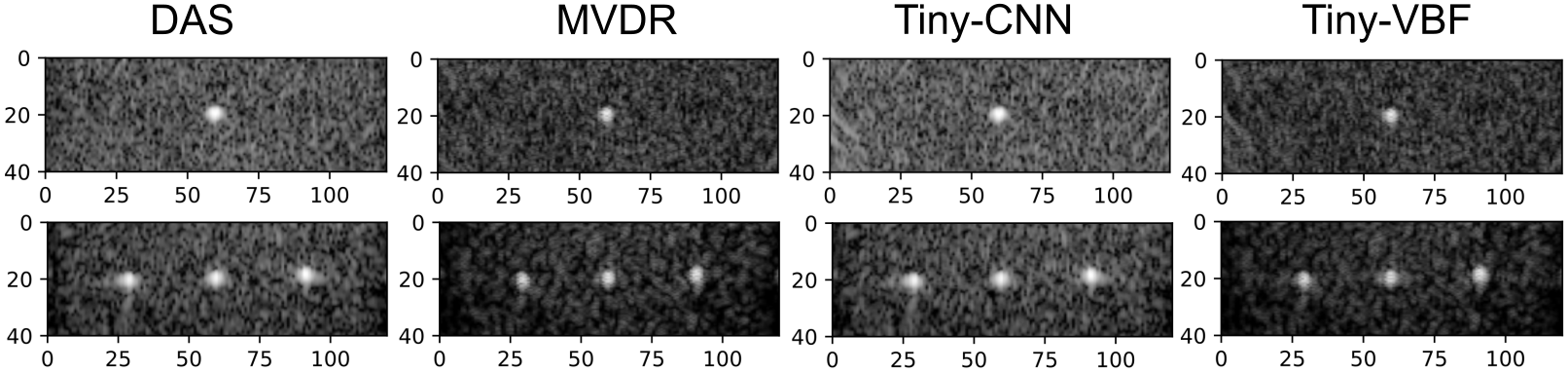}
\caption{B-mode images of in-vitro resolution-distortion dataset at depths of 14 mm and 33 mm}
\label{fig:my_label-13}
\end{figure}

 Similarly, the results over point targets in the experimental data is shown in Fig.~\ref{fig:my_label-13}. The performance of the Tiny-VBF model is consistently shown better than the DAS and Tiny-CNN in the experimental data. 

\par
\begin{figure}[h]
    \centering
  
  \subfloat[\label{1a}]{%
        \includegraphics[scale=0.1]{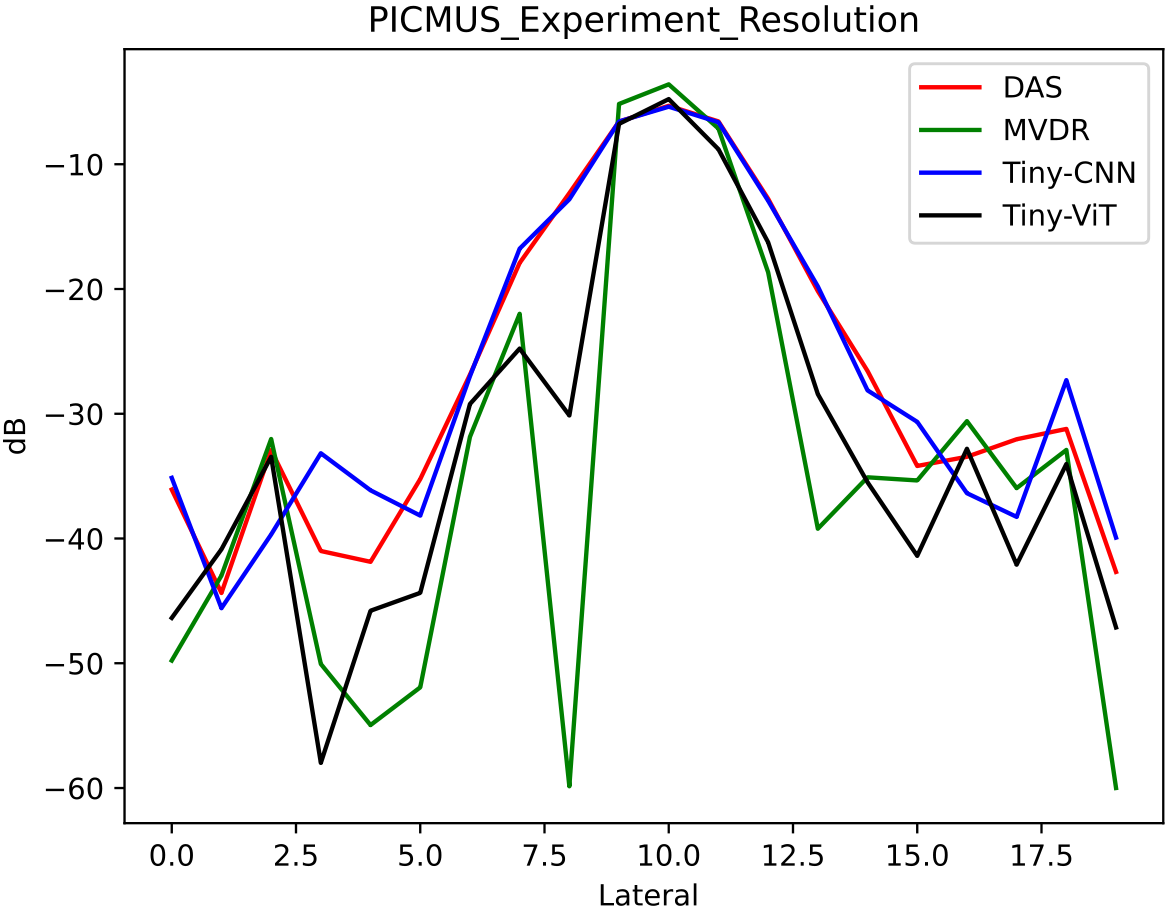}}
        \hspace{0.5mm}
  \subfloat[\label{1b}]{%
        \includegraphics[scale=0.1]{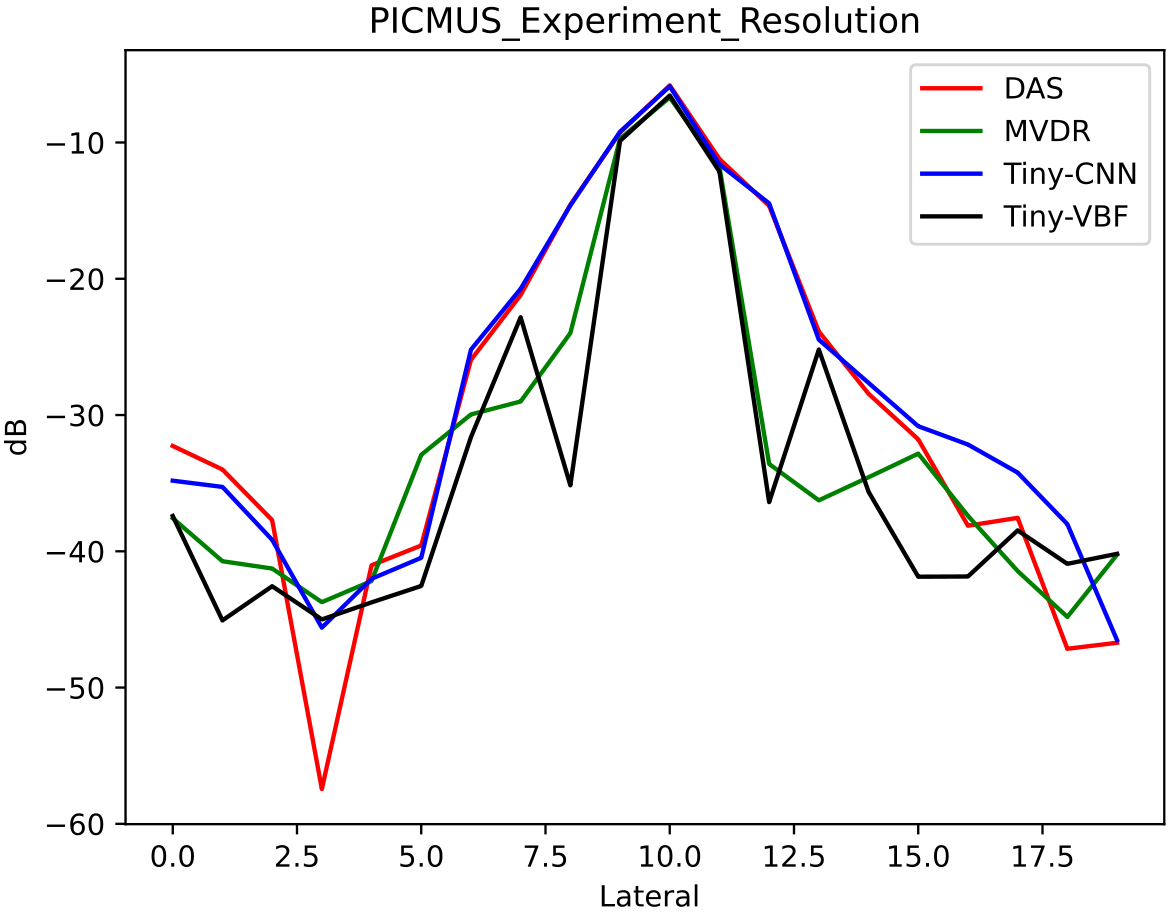}}
  \caption{Lateral point spread functions at (a) 14.01 mm and  (b) 32.79 mm where x-axis represents the lateral point position (pixel position) and y-axis represents the normalized amplitude values}
  \label{fig:my_label-14} 
\end{figure}

 Additionally, graphs in Fig.~\ref{fig:my_label-14} illustrate the lateral variation results from in-vitro data, demonstrating the heightened performance achieved with the Tiny-VBF model.

Moreover, Table~\ref{table-2} provides a quantitative comparison of axial and lateral resolution for both simulation and phantom data. It is evident that the proposed Tiny-VBF consistently surpasses the state-of-the-art DAS and Tiny-CNN in both cases. 

\begin{table}[!ht]
        \caption{Quantitatively analysis of axial and lateral resolution of Simulation and Phantom data}
        \centering
        \begin{tabular}{p{0.18\linewidth}p{0.08\linewidth}p{0.08\linewidth}p{0.08\linewidth}p{0.08\linewidth}}
            \hline
            \textbf{Beamformer} & \multicolumn{2}{c}{\textbf{Simulation (mm)}} &  \multicolumn{2}{c}{\textbf{Phantom (mm)}} \\
            
            & Axial & Lateral & Axial & Lateral  \\
            \hline
            DAS & 0.364 & 0.6 & 0.459 & 0.6 \\
            MVDR & \textbf{0.297} & \textbf{0.45} & 0.459 & \textbf{0.48} \\
            Tiny-CNN & 0.368 & 0.6 & 0.466 & 0.72 \\
            Tiny-VBF & \textbf{0.303} & \textbf{0.45} & \textbf{0.444} & \textbf{0.48} \\
            \hline 
        \end{tabular}
\label{table-2}
\end{table}

\par The inference time of Tiny-VBF model for each frame with frame size of \textit{368 x 128} is estimated around \textit{0.230 sec} on \textit{Intel Xeon 2vCPU@2.2GHz}. Whereas the inference time of Tiny-CNN, MVDR and CNN~\cite{unet} are approximately \textit{0.520 sec, 240 sec and 4 sec}, respectively.

\subsection{Evaluation on FPGA}
\par We quantized the Tiny-VBF model with different quantization levels and deployed it on FPGA. In addition, We used two hybrid quantization schemes as represented in Table~\ref{table-3}.

\begin{table}[!ht]
        \caption{Hybrid quantization bit-width representation}
        \centering
        \begin{tabular}{p{0.25\linewidth}p{0.15\linewidth}p{0.15\linewidth}}
            \hline
             & \textbf{Hybrid-1} & \textbf{Hybrid-2}\\
            \hline
            Weights & 8 bits & 8 bits  \\
            Softmax  & 24 bits & 24 bits\\
            Mul/Add ops & 20 bits & 16 bits  \\
            Intermediate output &  20 bits & 16 bits\\
            \hline 
        \end{tabular}
\label{table-3}
\end{table}

\begin{figure}[h]
\centering
\includegraphics[scale=0.2]{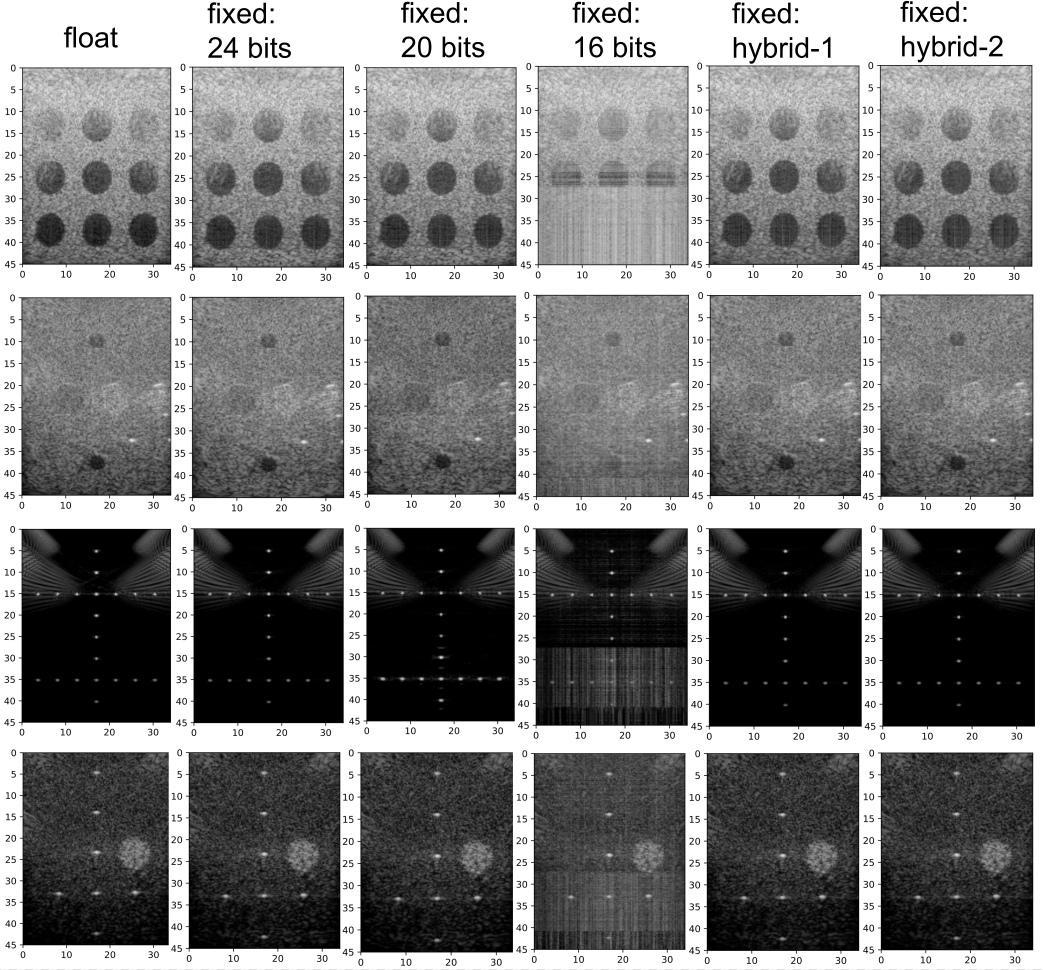}
\caption{B-mode images generated from FPGA}
\label{fig:my_label-15}
\end{figure}

The reconstructed images of simulation and phantom data, utilizing floating-point data and different quantization levels including hybrid quantizations, is shown in Fig.~\ref{fig:my_label-15}. We observed a significant degradation in image quality with the use of 16-bit quantization.

\par Table~\ref{table-4} displays the resolution of both simulation and phantom data with different quantization levels. The quantitative values were similar to original values in case of quantizations levels 24 bits, 20 bits and hybrid quantizations. Similarly, contrast metrics of both simulation and phantom data are shown in Table~\ref{table-5}. There is slight variation in the contrast values as compared to the floating point implementation.

\begin{table}[!ht]
        \caption{ Resolution metrics of Simulation and Phantom data on FPGA}
        \centering
        \begin{tabular}{p{0.18\linewidth}p{0.08\linewidth}p{0.08\linewidth}p{0.08\linewidth}p{0.08\linewidth}}
            \hline
            \textbf{Quantization} & \multicolumn{2}{c}{\textbf{Simulation (mm)}} &  \multicolumn{2}{c}{\textbf{Phantom (mm)}} \\
            \textbf{Bit-width} & Axial & Lateral & Axial & Lateral  \\
            \hline
            Float & \textbf{0.303} & \textbf{0.45} & 0.444 & \textbf{0.48} \\
            24 bits &  \textbf{0.303} & \textbf{0.45} & 0.444 & \textbf{0.48} \\
            20 bits & 0.310 & \textbf{0.45} & \textbf{0.421} & 0.54 \\
            Hybrid-1 & 0.309 & \textbf{0.45} & 0.429 & 0.54 \\
            Hybrid-2 & 0.309 & \textbf{0.45} & 0.429 & 0.54 \\
            \hline 
        \end{tabular}
\label{table-4}
\end{table}

\begin{table}[!ht]
        \caption{Contrast metrics of Simulation and Phantom data on FPGA}
        \centering
        \begin{tabular}{p{0.18\linewidth}p{0.08\linewidth}p{0.08\linewidth}p{0.08\linewidth}p{0.08\linewidth}p{0.08\linewidth}p{0.08\linewidth}}
            \hline
            \textbf{Quantization} & \multicolumn{3}{c}{\textbf{Simulation (dB)}} &  \multicolumn{3}{c}{\textbf{Phantom (dB)}} \\
            
            \textbf{Bit-widths} & CR & CNR & GCNR & CR & CNR & GCNR \\
            \hline
            Float & \textbf{14.89} & 1.75 & 0.74 & 12.20 & 1.39 & 0.67 \\
            24 bits & 14.07 & \textbf{1.84} & \textbf{0.75} & 13.0 & 1.22 & \textbf{0.69} \\
            20 bits & 14.30 & 1.45 & 0.73 & \textbf{13.05} & 1.22 & 0.67 \\
            Hybrid-1 & 13.34 & 1.74 & 0.73 & 12.72 & 1.37 & 0.68 \\
            Hybrid-2 & 13.26 & 1.75 & 0.72 & 12.62 & \textbf{1.40} & 0.67 \\
            \hline 
        \end{tabular}
\label{table-5}
\end{table}
\par Furthermore, Table~\ref{table-6} presents the resource utilization of the Tiny-VBF model at various quantization bit-widths. The quantization of parameters results in a significant reduction in resource requirements. In the case of Hybrid-2 quantization, the resource consumption is reduced by more than $50\%$ compared to its floating-point implementation.

\begin{table}[!ht]
        \caption{Resource utilization of Tiny-VBF model on FPGA}
        \centering
        \begin{tabular}{p{0.14\linewidth}p{0.08\linewidth}p{0.08\linewidth}p{0.08\linewidth}p{0.08\linewidth}p{0.08\linewidth}p{0.08\linewidth}}
            \hline
             \textbf{Resource} & \textbf{Float} & \textbf{24 bits}& \textbf{20 bits} & \textbf{16 bits}& \textbf{Hybrid-1} & \textbf{Hybrid-2}\\
            \hline
            LUT & 124935 & 88457 & 84594 & 59840 & 72415 & \textbf{61951} \\
            FF  & 91470 & 50454 & 43333 & 34920 & 38287 & \textbf{29105}\\
            BRAM & 161.5 & 158 & 156 & \textbf{82} & 150 & 110 \\
            DSP &  533 & 279 & 148 & 274 & \textbf{146} & 274 \\
            LUTRAM & 17589 & 11556 & 9442 & 6795 & 5352 & \textbf{5324}\\
            Power (W) & 4.489 & 4.369 & 4.174 & \textbf{3.989} & 4.229 & 4.174\\
            \hline 
        \end{tabular}
\label{table-6}
\end{table}

\section{Conclusion}
\par In this study, we introduced Tiny-VBF, a vision transformer-based image reconstruction for ultrasound imaging. The Tiny-VBF enhanced ultrasound beamforming quality when compared to the standard DAS method, in both in-silico and in-vitro data. The Tiny-VBF exhibit a minimal computational burden, requiring only 0.34 GOPs/Frame for a frame size of 368 x 128, as compared to the state-of-the-art deep learning models. We introduced an accelerator architecture and deployed our Tiny-VBF on the Zynq Ultrascale+ MPSoC ZCU104 FPGA. Our evaluation included network quantization and an assessment of the Tiny-VBF performance in terms of both quality and resource utilization. Notably, our hybrid quantization approach achieved a remarkable reduction of over $50\%$ in resource consumption without compromising image quality. To the best of our knowledge, this represents the first implementation of a vision transformer-based beamformer and the pioneering deployment of a deep learning-based beamformer on an FPGA.

\section*{Acknowledgment}
This work is supported in part by the NYUAD Center for Artificial Intelligence and Robotics, funded by Tamkeen under the NYUAD Research Institute Award CG010.

\balance

\vspace{12pt}

\end{document}